\documentclass[prl,showpacs,showkeys,twocolumn,notitlepage,10pt]{revtex4-1}
\usepackage{amsfonts,bbold,amsmath,amssymb,graphicx,epstopdf,verbatim,dsfont,color}
\usepackage[english]{babel}
\usepackage{natbib}

\newcommand{\tr}{{\rm Tr}}

\def \be{\begin{equation*}}
\def \ee{\end{equation*}}

\begin{document}

\title{Quantum generative model for sampling many-body spectral functions}
\author{Dries Sels}
\affiliation{Department of Physics, Harvard University, 17 Oxford st., Cambridge, MA 02138, USA}
\author{Eugene Demler}
\affiliation{Department of Physics, Harvard University, 17 Oxford st., Cambridge, MA 02138, USA}
\date{\today}

\begin{abstract}
Quantum phase estimation is at the heart of most quantum algorithms with exponential speedup. In this letter we demonstrate how to utilize it to compute the dynamical response functions of many-body quantum systems. Specifically, we design a circuit that acts as an efficient quantum generative model, providing samples out of the spectral function of high rank observables in polynomial time. This includes many experimentally relevant spectra such as the dynamic structure factor, the optical conductivity or the NMR spectrum. Experimental realization of the algorithm, apart from logarithmic overhead, requires doubling the number of qubits as compared to a simple analog simulator. 
\end{abstract}

\maketitle

\emph{Introduction}. --
Quantum computers possess the ability to solve problems that are intractable to classical ones. They can have superpolynomial speedup over the best known classical algorithm; so-called quantum supremacy~\citep{pres}. In order to demonstrate this supremacy attention has shifted from function problems such as implementing Shor's algorithm~\citep{shor94}, to sampling problems~\citep{google18}, as it appears that one does not need a full universal quantum computer to get quantum speedup~\citep{aaronson11,aaronson14,aaronson17}. For example, sampling from the output distributions of random quantum circuit, as recently performed on Google's Sycamore chip~\citep{googlesup}, classically requires a direct numerical simulation of the circuit, with exponential computational cost in the number of qubits.

While these random circuits have the virtue of being theoretically under control, meaning there is more confidence about the fact that they are hard to sample from than there is about factoring being hard, they are of limited practical use. They don't solve any problem other than providing evidence for quantum supremacy. Here, we trade some of the hardness for practical usefulness and provide a quantum circuit to obtain samples out of the spectral function of operators evolving under Hamiltonian dynamics in a many-body system. The problem essentially belongs to the class DQC1~\citep{knill98}, which is believed to be strictly smaller than BQP, while still containing classically intractable problems~\citep{Knill01,shor08}.

Spectroscopy is an important tool for characterising condensed matter and molecular systems. There is an entire plethora of techniques, each sensitive to different observables and in different parts of the energy spectrum. Many of those measurements can be formulated as a Fourier transform of some time dependent correlation function. Take for example, optical conductivity which probes the current-current correlations $\sigma(\omega)=\left< j(\omega) j(-\omega)\right>/i\omega$ or inelastic neutron scattering which measure the density-density correlations $ S_k (\omega)=\left< \rho_k(\omega) \rho_{-k}(-\omega) \right>$, etc.. Understanding the behavior of these correlation functions is one of the central goals in quantum many-body physics. For example, they allow to probe collective excitations of the system and to characterize universal dynamics close to quantum phase transitions~\citep{tolya05}. Furthermore, they can be a powerful tool for studying non-equilibrium dynamics~\citep{tolyaRMP,markus08,Aron_2008,hild14}. On a computational level, obtaining dynamical response functions is inherently difficult, as the coherent many-body dynamics induces large non-local correlations~\citep{datta07,parker18}. The exponential dimension of the underlying Hilbert space precludes exact methods and for large systems one typically has to rely on approximate methods such as density-matrix renormalization group (DMRG)~\citep{dmrg}, dynamical mean-field
theory (DMFT)~\citep{dmft}, semi-classical phase space methods~\citep{phase} or even time-dependent density functional theory (DFT). Each of these methods provides an accurate description for a particular class of problems but they all have limitations, e.g. long-range correlations are poorly captured by DMFT, and DMRG becomes intractable at late times or in higher dimensions. While much progress has been made in extending the regime of validity of all these methods, a universal solution to the quantum simulation problem does not exist as long as P$\neq$ NP~\citep{schuch08,aaronson11}.

Here we present a method to efficiently extract samples out of spectral functions using a quantum computer. The method requires a number of qubits that is proportional the volume of the system.  Under certain constraints -- which are met in most of the physically relevant situations -- the algorithm runs in polynomial time. We focus on the infinite temperature correlation function but extensions to finite and zero temperature are straightforward and briefly discussed at the end. 
Note that, even at infinite temperature, strong correlations can lead to many interesting phenomena such as anomalous diffusion~\citep{chertkov94,gopalakrishnan19}, impurity induced correlations~\citep{nagy17}, many-body localization~\citep{dima19} and excited state quantum phase transitions~\citep{caprio08}. Moreover, some spectroscopic techniques, such as electron spin resonance (ESR)~\citep{esr12} and nuclear magnetic spin resonance (NMR)~\citep{nmr05}, are naturally described by infinite temperature ensembles. 

The paper is structured in the following way. First, we discuss how to extract the spectrum by performing quantum phase estimation on a special purified state whose precise form depends on the operator of interest. It is this part of the algorithm which is responsible for the speedup. The fact that the entire operator content is represented in a single pure state eliminates the need to sample over all initial states, making it more efficient than performing analog Ramsey interferometry~\citep{knap13}. Second, we return to the question of preparing the required initial state and show that it does not degrade the speedup. We provide an explicit algorithm to construct the required states by postselection on an ancilla qubit.  Finally, we discuss extension to zero and finite temperature states. 

\emph{Quantum generative model}. --
\begin{figure*}[t!]
	\centering
	\includegraphics[width=0.8 \textwidth]{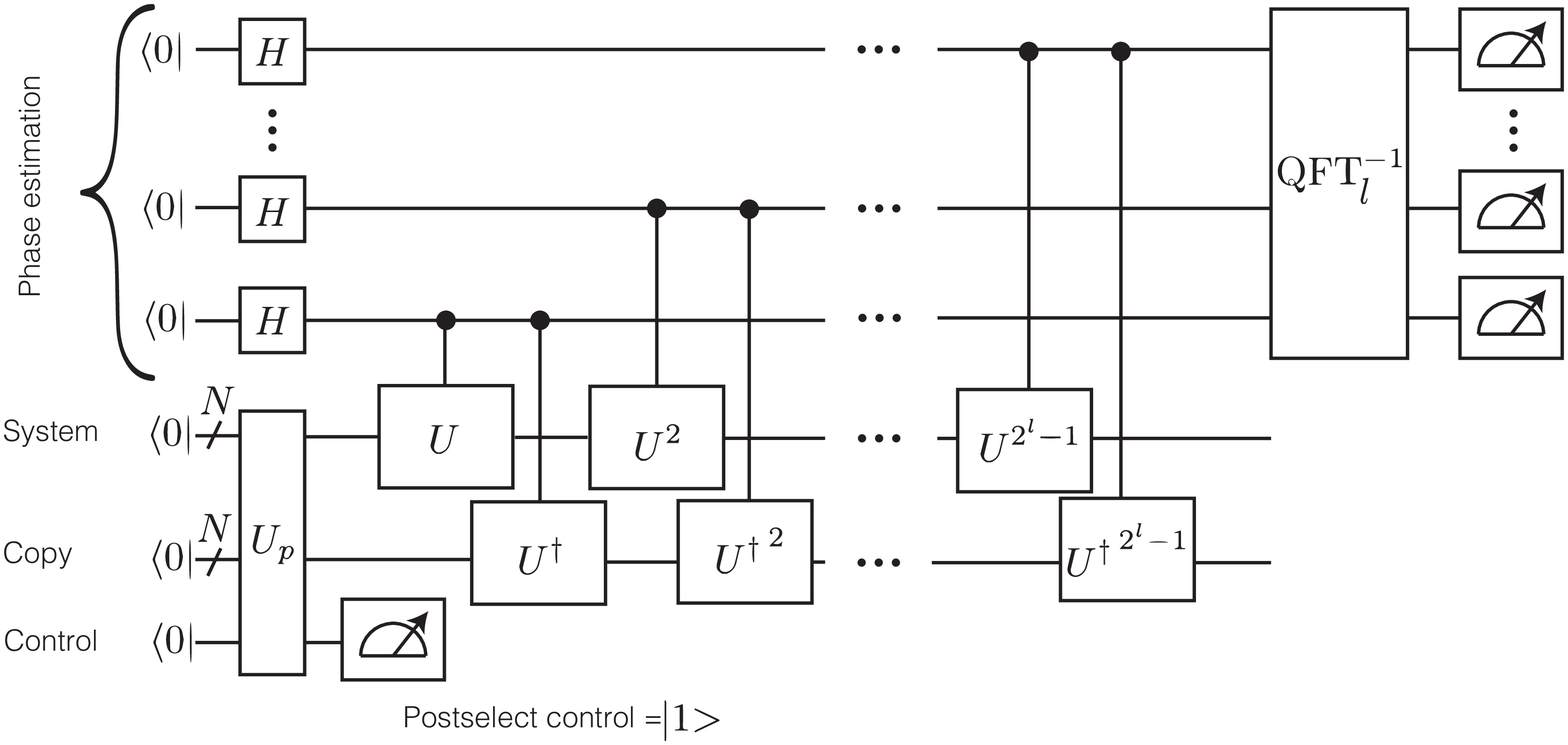}
	\caption{\textbf{Quantum circuit} Quantum phase estimation is performed on a purified operator. The purified state can be prepared by entangling two copies with an ancilla control qubit and postselecting the result on outcomes $\left| 1 \right>$, see Fig.\ref{fig:stateprep}. A phase difference between the two copies appears because each phase estimation bit propagates one copy according to $U$ and the other as $U^\dagger$. The output distribution after quantum Fourier transform is the spectral function.}
	\label{fig:diagram}
\end{figure*}
Consider the infinite temperature two-time correlation function:
\begin{equation}
S(t)=\frac{1}{\tr \left[\mathbb{ 1} \right]} \tr\left[  e^{iHt} O e^{-iHt} O \right],
\label{eq:Dt}
\end{equation}
of an operator $O$, undergoing dynamics according to Hamiltonian $H$. In particular, we are interested in obtaining samples out of its spectral function:
\begin{equation}
\Sigma_\gamma(\omega)= {\rm Re} \int_0^\infty {dt} e^{i\omega t-\gamma t} S(t),
\label{eq:spectrum}
\end{equation}
where $\gamma$ is the effective linewidth. 

We proceed by purifying~\citep{choi,stinespring,nielsen} a normalized version of the operator $O^2$, acting on the Hilbert space $\mathcal{H}$, into a pure state one an extended Hilbert space $\mathcal{H} \otimes \mathcal{H}$: 
\begin{equation}
\left| O \right> = \mathcal{N}^{-1/2}\sum_{i=1}^{2^N} O_i \left| i \right> \otimes \left| i \right>,
\label{eq:purO}
\end{equation}
where $O_i$ and $\left|i \right>$ are the eigenvalues  and eigenvectors of $O$ respectively. The normalization is simply $\mathcal{N}= \tr \,O^2 $. Next, we perform quantum phase estimation on the unitary which propagates one of the two copies with the actual Hamiltonian $H$ and the other copy with $-H$, such that a phase difference accumulates between the copies over time. If we denote 
\begin{equation}
H=\sum_{n=1}^{2^N} \epsilon_n \left| E_n \right> \left< E_n \right|,
\end{equation}
 then quantum phase estimation on the state $\left| O \right>$, results in the state:
\begin{eqnarray}
\left| \Psi \right> &=& \sum_{n,m=1}^{2^N} 2^{-l/2}\sum_{x=0}^{2^l-1}  c_{n,m} e^{i \Delta(\epsilon_n-\epsilon_m)x }\left| E_n \right> \otimes \left| E_m \right> \otimes \left| x \right>, \nonumber   \\ 
&& {\rm with }\quad c_{n,m}= \frac{\sum_i \left< E_n | i\right>\left< E_m | i\right>O_i}{\sqrt{\mathcal{N}}}.
\end{eqnarray}
Here $l$ denotes the number of ancilla qubits used to perform the quantum phase estimation and $\left| x \right>$ denotes the computational basis state of the ancilla given by the binary representation of $x$, e.g. $x=2$ implies $\left|0 \cdots 0 1 0\right>$. Finally $\Delta$ denotes the effective time for which the control (phase estimation) qubit is coupled to the system. See Fig.~\ref{fig:diagram} for a circuit representation. 
Performing an \emph{inverse quantum Fourier transform}~\citep{nielsen,shor94} on this state one arrives at:
\begin{eqnarray}
&&\left| \Psi_{ \rm QFT} \right> = \sum_{n,m=1}^{2^N}\sum_{k=0}^{2^l-1} c_{n,m} A_{n,m}^k \left| E_n \right> \otimes \left| E_m \right> \otimes \left| k \right>, \nonumber \\
&& {\rm with}\,A_{n,m}^k=\frac{1}{2^l}\sum_{x=0}^{2^l-1}  \exp{\left[i \frac{2\pi}{2^l} \left( \frac{\Delta2^l}{2\pi}(\epsilon_n-\epsilon_m)-k \right) x \right]}. \nonumber 
\end{eqnarray}
Finally a measurement is performed on the phase estimation qubits in the computational basis, see Fig.~\ref{fig:diagram}. The probability to find the control bits in state $\left| f \right>$ is simply given by:
\begin{equation}
P(f)=\sum_{n,m=1}^{2^N} |c_{n,m}|^2 \left| A_{n,m}^f \right|^2
\label{eq:Pf}
\end{equation}
Assuming time-reversal symmetry of the Hamiltonian $H$ and operator $O$, one finds
\begin{equation}
|c_{n,m}|^2=\frac{|\left< E_n \right| O \left| E_m \right>|^2}{\tr\left[ O^2 \right]}.
\end{equation}
This is exactly the (normalized) golden rule transition rate between energy eigenstates. Moreover, the second part in expression~\eqref{eq:Pf} is a function that concentrates around $f=\Delta 2^l (\epsilon_n -\epsilon_m)/2\pi$, i.e.
\begin{eqnarray}
\left| A_{n,m}^f \right|^2 &=&  \frac{1}{4^l}\frac{ \sin^2\left[ \pi \left( \frac{\Delta2^l}{2\pi}(\epsilon_n-\epsilon_m)-f \right) \right]  }{ \sin^2\left[ \frac{\pi}{2^l} \left( \frac{\Delta2^l}{2\pi}(\epsilon_n-\epsilon_m)-f \right) \right] } \nonumber \\
 &\geq& {\rm sinc}^2\left[ \frac{\Delta2^l}{2\pi}(\epsilon_n-\epsilon_m)-f  \right],  
\end{eqnarray}
with ${\rm sinc}(x)=\sin(\pi x)/\pi x$. Consequently, for carefully chosen parameters the output distribution of the phase estimation qubits is exactly the desired spectral function: $P(f)\sim \Sigma_\gamma(\omega \Delta 2^l/2\pi)$. A proper spectral measurement requires: 
\begin{equation}
\frac{1}{\gamma}\leq \frac{\Delta 2^l}{2\pi} \leq \frac{2^l-1}{\omega_{max}}.
\end{equation}
The first inequality expresses the fact that one at least needs to resolve frequencies at a better level than the effective linewidth $\gamma$. The second simply states that a minimal amount of bits are required to resolve the bandwidth $\omega_{max}=\max (\epsilon_n -\epsilon_m)$. With $l$ bits, there are $2^l$ configurations while the number of distinguishable peaks is $\sim  \omega_{max}/\gamma$, consequently the number of bits should scale like
\begin{equation}
l \propto \log \frac{\omega_{max}}{\gamma}.
\end{equation}
For any problem in which the bandwidth scales polynomial with the system size $N$ and for which the linewidth decreases algebraically in the system size, the number of phase estimation qubits scales logarithmically in $N$. Note that this is the case in almost all physically relevant situations. First, for local models, the bandwidth simply scales linearly in the system size and even systems with all-to-all interactions only have quadratic scaling of the bandwidth with system size. Second, with a few exceptions, one is typically only interested in studying the behavior of the system for a time $T$ which is polynomial in the system size. In that case, an algebraically small linewidth should be sufficient. Finally,  $\Delta\approx 2\pi/\omega_{max}$, which is not unreasonable for polynomial bandwidth. 
Note that it appears that we need $O(l)$ gates to apply the controlled unitaries in~Fig.~\ref{fig:diagram}, however all those gates commute and can in principle be done in parallel. The last gate can nonetheless not be implemented in the same physical time as the first, while the first gate only takes a time $O(\Delta)$ the last gate requires a time of $O(\gamma)$.  A standard implementation of QFT takes $O(l^2)$ gates~\citep{nielsen}, but more sophisticated versions only require $O(l\log l)$ gates~\citep{hales00}. Therefore the computational time scales is at worst $O(\gamma^{-1}+l^2)$ or $O(\omega_{max}/\gamma+l^2)$ if one has to decompose the Hamiltonian $H$ into two-qubit gates

\emph{Initial state preparation}. --
The efficiency of the above procedure hinges on the ability to prepare the initial state $\left| O \right>$. We provide an explicit probabilistic method to prepare $\left| O \right>$ out of a product state by postselecting on the measurement outcome of an ancilla qubit. First of all note that, if operator $O$ would be of low rank, the above procedure would be superfluous. In the latter case, one could simply extract the two-point function~\eqref{eq:Dt} by evolving each of the eigenvectors of $O$. Only ${\rm rk}(O)$ states would have to be propagated, so it can be done in polynomial time as long as the rank is polynomial in the system size. We wish to obtain a method for operators that have no, or only small, rank deficiency. 

\begin{figure}[t!]
	\includegraphics[width=0.45 \textwidth]{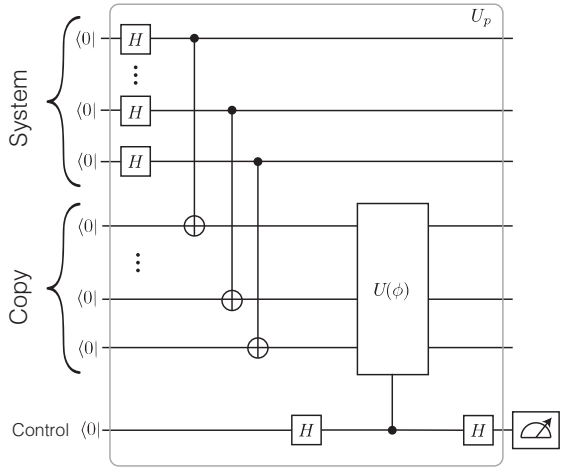}
	\caption{\textbf{State preparation scheme} An initial entangled pair states is created between two $N$-qubit registers. Next, one of the two copies is connected to an ancilla control qubit, which is placed in an equal superposition of z-states, i.e. both are evolved for some time $\phi$ under the Hamiltonian, $H=O (\sigma^z_a+1)/2$. Performing another Hadamard gate on the ancilla and postselecting the outcome on $\left| 1 \right>$, the entangled pair state will be transformed into the desired $\left| O \right>$ state. The success probability is of the procedure is determined by the ratio of the typical value of $O^2$ to its maximal value $O_{max}^2$.}
	\label{fig:stateprep}
\end{figure}

Let us start by preparing a maximally entangled pair state
\begin{equation}
\left| \psi_{\rm EP} \right>=\sum_{i=1}^{2^N} \frac{1}{\sqrt{2^N}} \left| z_i \right> \left| z_i \right>,
\label{eq: }
\end{equation}
and try to project the system to the desired state $\left| O\right>$; note that $\left| O \right> \propto O \otimes \mathbb{1} \left| \psi_{\rm EP} \right>$.  The creation of the entangled pair state $\left| \psi_{\rm EP} \right>$ is relatively easy, it's simply a product state of Bell pairs between the system and its copy. It can be constructed out of a product state in constant time, see Fig.~\ref{fig:stateprep}. 

A single control qubit can now be used to apply a controlled unitary rotation, with the action on the system being:
\begin{equation}
U(\phi)= \exp (i \phi O )\otimes \mathbb{1}
\end{equation}
By applying a Hadamard gate on the control bit before and after $U$, the combined state becomes: 
\begin{equation}
\left| \psi \right> =\frac{1}{2} (1+U(\phi))  \left| \psi_{\rm EP} \right> \left| 0\right>+\frac{1}{2}(1-U(\phi))   \left| \psi_{\rm EP} \right> \left| 1 \right>.
\end{equation}
Measuring the control qubit in the computational basis, one finds it in the $\left| 1 \right>$ state with probability 
\begin{equation}
P_1(\phi)=\left< \psi_{\rm EP} \right| \sin^2 \left( \phi O/2 \right) \left| \psi_{\rm EP} \right>.
\label{eq:P1}
\end{equation}
At the same time, the fidelity between the target state $\left| O \right>$ and the postselected state $\left |\psi_1\right>$ becomes
\begin{equation}
F(\phi)= \vert \left< O | \psi_1 \right>\vert^2=\frac{\vert \left<  O U(\phi)  \right>\vert^2 }{\left< O^2\right> \left< 4\sin^2 \left( \phi O/2 \right) \right>},
\label{eq:Fid}
\end{equation}
where the averages are in the infinite temperature state $\left| \psi_{EP} \right>$, without loss of generality we assumed $O$ to be traceless. The fidelity tends to 1 when $\phi \rightarrow 0$, however, at the same time the acceptance probability also goes down. To be efficient, we need to achieve a fidelity $F=1-\epsilon$ with a probability that is at worst algebraically small in $N$. For sufficiently small $\phi$, we find
\begin{equation}
P_1= \frac{\phi^2}{4} \left< O^2\right>+O(\phi^4),
\end{equation}
while
\begin{equation}
F=1-\frac{\phi^2}{4} \left(\frac{\left< O^4\right>}{\left< O^2 \right>} - \frac{\left< O^3 \right>^2}{\left< O^2 \right>^2} \right)+O(\phi^4).
\end{equation}
Consequently, as long as higher order contributions can be neglected, one gets a fidelity better than $1-\epsilon$ by setting $\phi^2=\epsilon \left< O^2 \right>/ \left< O^4 \right>$, resulting in success with probability
\begin{equation}
P_1 \approx \epsilon \frac{\left< O^2 \right>^2}{\left< O^4 \right>} \geq \frac{\left< O^2 \right>}{O_{max}^2}\geq \epsilon \frac{{\rm rk}(O)}{2^N} \left( \frac{O_{min}}{O_{max}}\right)^2
\end{equation}
where $O_{max}$ is the largest singular value of $O$ and $O_{min}$ is the smallest non-zero singular value.  For most physical observables, such as those comprised of sums of local terms, the fourth moment simply scales as the square of the second, i.e. $\left< O^4 \right> \propto \left< O^2 \right>^2$. Hence, for all those observables the state can be prepared in a constant time of $O(1/\epsilon)$. Additionally, it's sufficient that the operator only has polynomial rank deficiency and polynomial scale separation between its smallest and largest singular value, to be able to generate the state in polynomial time.
\begin{figure}[t!]
	\includegraphics[width=0.45 \textwidth]{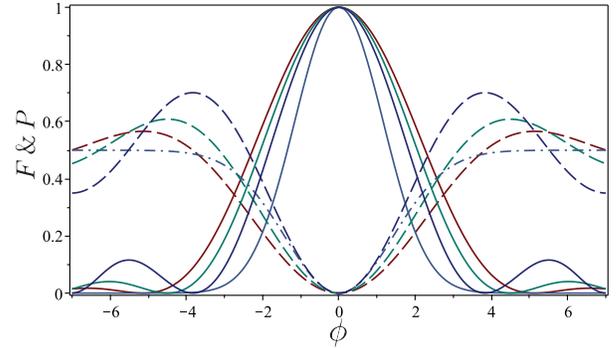}
	\caption{\textbf{Preparation efficiency} Fidelity between the post-selected state $\left| \psi_1 \right>$ and the target state $\left| O \right>$ decays with the rotation angle $\phi$ of the controlled unitary rotation $U(\phi)$ (full lines). Similarly, the success probability increases from $0$ to $1/2$ when the angle increases (dashed lines). Different curves show expressions~\ref{eq:Fid} and~\ref{eq:P1} for different eigenvalue distributions of $O$, i.e. results are shown for Wigner semicircle, uniform, arcsine and Gaussian eigenvalue distributions. Each of these distributions has a success probability $P=c (1-F)$ in a broad region of $\phi$'s around zero. The constant $c=O(1)$ for all distributions, i.e. $1/2$, $5/9$, $2/3$ and $1/3$ for the semicircle, uniform, arcsine and Gaussian respectively.}
	\label{fig:distributions}
\end{figure}

\emph{Discussion}. --
Even at infinite temperature, the dynamical properties of operators evolving under a many-body Hamiltonian are theoretically interesting. In particular their spectral function provides information about the universal behavior of the system~\citep{luca16,parker18}. Both, the high and low frequency behavior of the spectral function is universal and while the former gives insight into the Lyapunov exponent of the operator, the latter provides information about the diffusion constant. 

Apart from theoretical interest, there is at least one relevant problem which is effectively at infinite temperature, namely nuclear magnetic resonance (NMR) spectroscopy. In NMR one measures the response of the nuclear spins of system placed in high magnetic field to an external drive, i.e. $O=\sum_{i=1}^N \sigma^z_i$. These systems are not isolated from the environment, yet have relatively long but finite coherence time. As a consequence, $\gamma$ is finite and $P_1\sim 1/\epsilon$ and the entire  algorithm runs in a time $t=O(\epsilon^{-1}+\gamma^{-1}+l^2)$, which to leading order in $N$ is $\log^2 N$. 

Finally, it's interesting to extend the present results to finite and zero temperature. There was nothing specific about the phase estimation scheme, one simply has to purify a different operator. At zero temperature, expression~\eqref{eq:purO} has to be replaced with 
\begin{equation}
\left| O \right>_0  =\frac{1}{\sqrt{\left< O^2 \right>}} O \left| \psi_0 \right> \otimes \left| \psi_0 \right>.
\end{equation}
If the ground state $ \left| \psi_0 \right>$ can be efficiently prepared, the preparation of $\left| O \right>_0$ might continue as before, with a similar success rate. One only has to replace the expectation in~\eqref{eq:P1} with ground state expectation values.  Consequently, for local observables we still expected $P_1=O(1/\epsilon)$. Note that the state $\left| O \right>_0$ is a product state between the system and the copy, hence the copy only serves as a reference for the phase. If one knows the ground state energy, or doesn't care about shifts in the spectrum, one can eliminate the copy entirely. Finally, in order to sample from any finite temperature spectral function, one simply has to replace the maximally entangled pair state $\left| \psi_{\rm EP} \right>$ with the less entangled purification of a Gibbs state:
\begin{equation}
\left| \psi_\beta \right> = Z^{-1/2}\sum_n e^{-\beta \epsilon_n/2} \left| E_n \right> \otimes \left| E_n \right>,
\end{equation}
such that $\left| O \right>_\beta \propto O \left| \psi_\beta \right>$; it clearly tends to the zero and infinite temperature state for large and small $\beta$ respectively. If the purified Gibbs state can be made efficiently, the algorithm is just as efficient as before. Whether or not this is possible, depends entirely on the problem at hand, i.e. a QMA-complete problem might have been embedded in the Hamiltonian, implying it can not take less then exponential time. On the other hand, many physically relevant problems are expected to be less hard. At zero temperature, one can imagine an adiabatic preparation procedure and as long as there is no exponential gap closing this should work in polynomial time. For $\left| \psi_\beta \right>$, one might have to resort to numerical optimal control methods to find efficient state preparation schemes~\citep{crab15,bukov18}.

\emph{Acknowledgements}. -- DS acknowledges support from the FWO as post-doctoral fellow of the Research Foundation -- Flanders. ED acknowledges support from the Harvard-MIT CUA, AFOSR-MURI: Photonic Quantum Matter (award FA95501610323), DARPA DRINQS program (award D18AC00014), AFOSR-MURI: Quantum Phases of Matter (grant FA9550-14-1-0035).The authors acknowledge discussion with H. Pichler, M. Lukin, X. Gao, O. Demler, H. Dashti, S. Mora. 

\bibliography{genmodbib}

\end{document}